\documentclass{epl2}

\title{Shear-thickening induced by Faraday waves in dilute wormlike micelles}
\shorttitle{Shear-thickening induced by Faraday waves}
\author{P. Ballesta\thanks{E-mail: \email{ballesta@crpp-bordeaux.cnrs.fr}} \and S. Manneville}
\institute{Centre de Recherche Paul Pascal, Avenue Schweitzer, 33600 Pessac, FRANCE}


\abstract{
We investigate the effect of surface waves generated by the Faraday instability on a shear-thickening surfactant solution under vertical vibrations. We show that a prolonged oscillation of the surface above the instability onset leads to an increase of the fluid viscosity. This phenomenon is evidenced by comparing the time needed for the instability to develop in the fluid at rest and once the fluid has been shaken above onset: pre-shaking may delay the instability by two orders of magnitude. A simple model based on a time-dependent viscosity is proposed which accounts quantitatively for the experimental observations.}

\begin{document}

\maketitle

\section{Introduction} 
In the last few decades, the coupling between microstructure and flow in complex fluids has been the subject of intense research effort \cite{Larson:1999}. In most cases the flow is a simple shear flow induced by a rheometer. Yet, another way to induce a controlled flow is by the means of hydrodynamic instabilities. In this paper the effect of the well-known Faraday instability, by which the initially flat surface of a fluid layer submitted to vertical vibrations gives way to surface waves with characteristic wave number $k_c$ above some critical acceleration $a_c$ \cite{Faraday:1831}, is tested on a shear-thickening dilute solution of wormlike micelles \cite{Liu:1996}. Recently the Faraday instability in complex fluids has been the subject of various studies \cite{Raynal:1998,Wagner:1999,Kumar:1999,Huber:2005,Ballesta:2005,Lioubashevski:1999,Merkt:2004,Kityk:2006}. In most of these previous works \cite{Raynal:1998,Wagner:1999,Kumar:1999,Huber:2005,Ballesta:2005}, the authors compared the behaviour of a complex fluid to that of a Newtonian one close to the instability threshold. Differences were explained by invoking the linear viscoelastic properties of the samples. Far above the instability threshold, a stronger coupling between the flow induced by the instability and the complex fluid microstructure is expected. For instance, in clay suspensions, hysteresis and high 
amplitude fingerlike structures were shown to be linked to shear-thinning~\cite{Lioubashevski:1999}. In shear-thickening colloidal suspensions, external perturbations may grow and get stabilized as ``persistent holes'' across the fluid surface \cite{Merkt:2004}. In a dilute polymer solution, the flow--microstructure coupling can delay the transition to disordered states~\cite{Kityk:2006}. The present work shows that, far above the instability threshold, the surface waves become large enough to induce shear-thickening in dilute wormlike micelles. In the following we first present the complex fluid under study and the experimental setup. Then measurements of the {\it set-up time} of the Faraday instability, defined as the time needed for the surface waves to become large enough to be detected by our experimental setup, are presented. We show that the set-up time may increase by two orders of magnitude when the fluid is pre-shaken above the instability threshold. Finally a model is proposed to account for such observations, based on a time-dependence of the viscosity induced by the surface waves.

\section{System under study and rheological measurements} 
\begin{figure}[t]
\onefigure{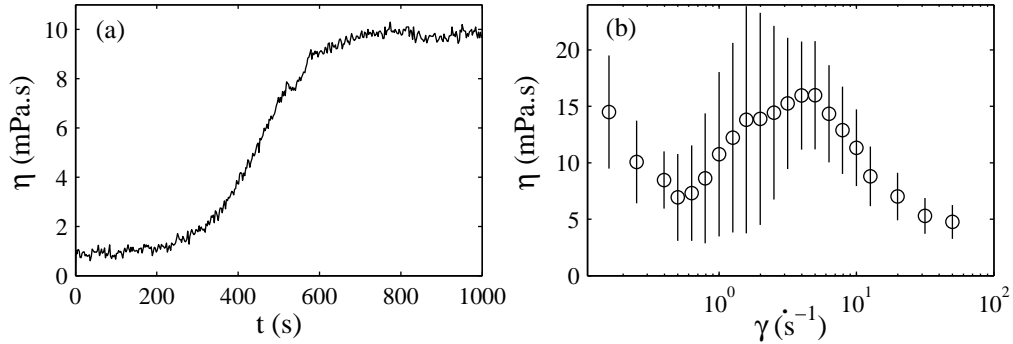}
\caption{(a) Temporal evolution of the viscosity of a $250$~ppm CTAB-NaSal solution for $\dot\gamma=5$~s$^{-1}$ at $T=20^{\circ}$C. (b) Viscosity $\eta$ as a function of the shear rate $\dot\gamma$. See text for experimental details.}
\label{f.1}
\end{figure}

Our working fluid is a dilute solution of wormlike micelles made of CTAB--NaSal in water at a very low concentration $c_{CTAB}=250$~ppm by weight and the molar ratio of salt to surfactant is 1:1. This wormlike micelle solution in the dilute regime is known to display a shear-thickening behaviour at typical shear rates $\dot\gamma=1$--10~s$^{-1}$~\cite{Liu:1996}. Such shear-thickening has also been observed in a wide range of other wormlike micelle solutions and appears as a common feature of dilute systems \cite{Boltenhagen:1997,Hu:1998,Berret:2002}. In order to characterise our low-viscosity micellar system, rheological measurements are performed in a double Couette geometry (rotor radii 28 and 34~mm, stator radii 29.5 and 32~mm) thermostated at $T=20^{\circ}$C using a strain-controlled Rheometrics RFSII rheometer. Figure~\ref{f.1}(a) shows the dynamic viscosity $\eta$ as a function of time when a shear rate of 5~s$^{-1}$ is imposed at time $t=0$. After an induction time of about 200~s, $\eta(t)$ increases by a factor of 10 within a characteristic time of about 500~s. This temporal evolution of the viscosity is comparable to that reported in the same system by Liu and Pine \cite{Liu:1996}.
Next a constitutive curve $\eta$ vs $\dot\gamma$ is obtained by submitting our fluid to various shear rates $\dot\gamma$ for $1000$~s and by averaging $\eta(t)$ over the last $100$~s. Such a procedure is repeated three times. The viscosity shown in fig.~\ref{f.1}(b) is the mean of these three measurements and the error bars are their standard deviation. It is clearly seen that between $\dot\gamma=0.5$ and 5~s$^{-1}$, the viscosity increases with increasing shear rate up to $\eta=16$~mPa.s. Such a shear-thickening phenomenon was attributed to the formation of a shear-induced gel-like structure \cite{Rehage:1986,Barentin:2001}, which was confirmed by light scattering experiments \cite{Boltenhagen:1997}. At greater shear rates, the fluid becomes shear-thinning as expected for a gel that gets progressively broken down by shear. The curve $\eta(\dot\gamma)$ of fig.~\ref{f.1}(b) is only qualitatively similar to that of ref.~\cite{Liu:1996}. In our opinion, the large error bars shown by our data in the shear-thickening regime illustrate the variability in both the final value of the viscosity and in the induction and growth times of $\eta(t)$, as well as the sensitivity of viscosity measurements to filling conditions and contamination in such dilute samples. 

\section{Experimental setup for the Faraday instability} 
Our experimental setup has been described in ref.~\cite{Ballesta:2005}. A cylindrical cell of depth $10$~mm and diameter $60$~mm filled with our fluid and thermostated at $T=20\pm0.1^{\circ}$C is submitted to vertical vibrations of frequency $f$ by an electromagnetic shaker (Ling Dynamic Systems V406). The fluid layer is lit from above so that the surface is seen bright only when it makes an angle greater than $1^{\circ}$ with the horizontal plane. Close to the instability onset, small amplitude surface patterns can be approximated by a superposition of sinusoidal waves of wave number $k_c$ and amplitude $\xi$. With our setup, surface waves are detected only when $\xi>\xi_d=\tan(\pi/180)/k_c$. In our experiments, $k_c\simeq1000$~m$^{-1}$ so that $\xi_d\simeq 20$~$\mu$m.

In a first step, measurements of the critical acceleration $a_c$ and wave number $k_c$ (not shown) were performed. As reported earlier by Raynal {\it et al.} \cite{Raynal:1998}, the viscoelastic behaviour of the solution leads to $a_c$ values lower than those of a Newtonian fluid with the same zero-shear viscosity while $k_c$ follows the same dispersion relation. Besides, in our system, the response of the surface is always subharmonic ({\it i.e.} at $f/2$) over the frequency range $f=20$--200~Hz and $a_c$ measurements do not reveal any significant hysteresis. Two elements may explain the absence of any signature of shear-thickening in these measurements. First the time needed to measure $a_c$ and $k_c$ at a given frequency ($\sim 3$~min) is smaller than the characteristic time for shear-thickening ($\sim 10$~min). Second although the equivalent shear rate, estimated as $\dot\gamma=\pi f \xi_d k_c$ close to instability onset \cite{Kumar:1999}, falls into the shear-thickening regime (one finds $\dot\gamma~\simeq 6$~s$^{-1}$ for $f=100$~Hz), the wave amplitude may still be too small to induce any modification of the fluid microstructure because the flow involved in the Faraday experiment is both oscillating and elongational and thus strongly differs from the continuous shear used in rheological experiments. Therefore, in order to induce shear-thickening through the Faraday instability, the experiments detailed below were performed on longer time scales and at accelerations far above the instability threshold where the wave amplitude is larger.

\section{Experimental results} 
\begin{figure}[t]
\onefigure{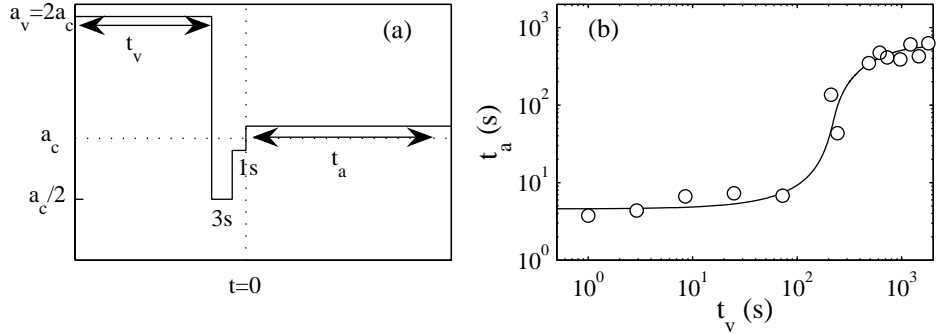}
\caption{(a) Experimental procedure. (b) Set-up time $t_a$ versus the pre-shaking duration $t_v$ at $a_v=2a_c$ and $f=100$~Hz. The continuous line is the result of the best fit by eqs.~(\ref{tab.1}) and~(\ref{tab.2}) with $\tau=461$~s and $\delta_{\infty}=0.149$. }
\label{f.2}
\end{figure}

To evidence a signature of shear-thickening, we measure the set-up time $t_a$ of the instability in our fluid once it has been shaken at an acceleration $a_v$ above onset for a duration $t_v$. The exact experimental procedure is sketched in fig.~\ref{f.2}(a). The fluid layer is first vibrated at frequency $f$ and $a_v=2a_c$ for a duration $t_v$. Such a ``pre-shaking'' is followed by a short period when the instability is stopped ($a=a_c/2$ for $3$~s then $a=0.9a_c$ for $1$~s) and the surface wave completely disappears. Finally the acceleration is set to $a=(1+\epsilon)a_c$ and the set-up time, {\it i.e.} the time needed for the surface wave to be detected again by our experimental set-up, is measured. This method is preferred to a direct measurement of $a_c$ because the time needed to precisely measure $a_c$ may vary from $1$ to $5$~min and therefore is not suitable for a time-dependent microstructure. In all cases, two successive measurements of $t_a$ are separated by at least ten minutes, in order to make sure that our fluid is not perturbed by previous excitation. 

In a first experiment, $t_v$ is varied from $1$ to $1800$~s for $f=100$~Hz and $\epsilon=0.05$. Figure~\ref{f.2}(b) shows that $t_a$ strongly increases with increasing $t_v$, up to two orders of magnitude. More precisely, $t_a$ remains roughly constant up to $t_v\simeq 200$~s which corresponds to the induction time mentioned above in fig.~\ref{f.1}(a). $t_a$ then saturates to about 600~s which again is consistent with the characteristic growth time of the shear-induced structure. We interpret this remarkable effect on the set-up time as a consequence of shear-thickening induced by pre-shaking above onset.

\begin{figure}[t]
\onefigure{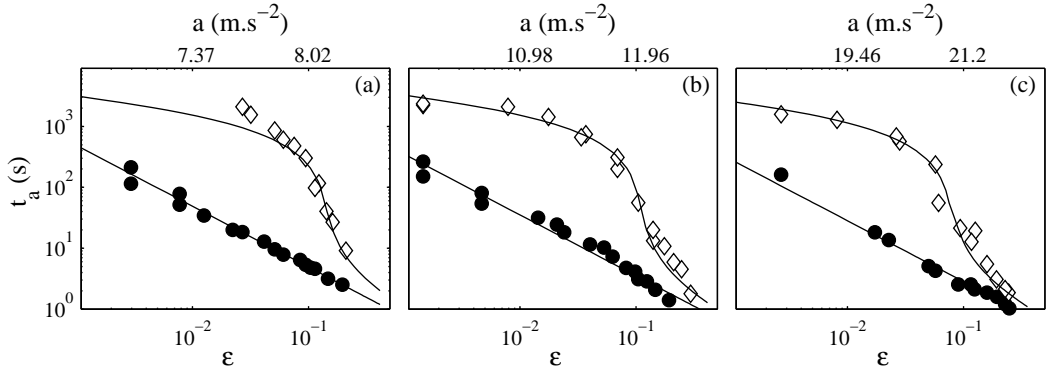}
\caption{Set-up time $t_a$ versus $a$ (top scale) and $\epsilon$ (bottom scale) for the fluid at rest ($\bullet$) and for the pre-shaken fluid ($\diamond $). The continuous line is the best fit by eqs.~(\ref{taf.1}) and~(\ref{taf.2}). The vibration frequency is (a) $f=60$~Hz, (b) $f=85$~Hz, and (c) $f=135$~Hz. The fit parameters are given in table~\ref{t.1}.}
\label{f.3}
\end{figure}

In a second experiment, $t_a$ is measured after pre-shaking at fixed $a_v=2a_c$ and $t_v=600$~s for various final accelerations $a$ above $a_c$. The results are shown in fig.~\ref{f.3} and compared to the set-up time measured in the fluid initially at rest ({\it i.e.} without pre-shaking) for three different frequencies. Let us first discuss the results obtained in the fluid at rest. As expected, one finds $t_a\propto 1/\epsilon$, where $\epsilon = a/a_c-1$ is the dimensionless distance to the threshold. Indeed, in general, as long as the amplitude $\xi$ of the surface wave is small enough, $\xi(t)$ follows the equation \cite{Douady:1990}: 
\begin{equation}
\label{Ampl.0}
\tau_g \frac{\hbox{\rm d}\xi}{\hbox{\rm d}t} = \epsilon \xi ,
\end{equation}
where $\epsilon/\tau_g$ is the growth rate, so that for a perturbation of initial amplitude $\xi_0$, the set-up time $t_a$ defined by $\xi(t_a)=\xi_d$ is given by:
\begin{equation}
\label{ta.2}
t_a = T_0\frac {a_c}{a-a_c},
\end{equation}
where $T_0=\tau_g\ln(\xi_d/\xi_0)$ is the set-up time for $\epsilon=1$ (i.e. $a=2a_c$). As shown in fig.~\ref{f.3}, eq.~(\ref{ta.2}) perfectly matches the experimental data in the fluid at rest and allows us to determine precisely $a_c$ hereafter noted $a_0$ for the sake of clarity (to within less than $1\%$) and $T_0$ (to about $2\%$). The best fit parameters are given in table~\ref{t.1}.

Let us now turn to the set-up time in the pre-shaken fluid. As seen in fig.~\ref{f.3}, for a given $\epsilon$, $t_a$ is much larger (sometimes a hundred times larger) for the pre-shaken fluid than for the fluid at rest. In the following, we argue that this behaviour is linked to an increase of the fluid viscosity induced by the surface waves when the fluid is shaken at $a_v=2a_0$. After the instability is stopped and the acceleration is set to its final value, the viscosity tends to relax towards that of the fluid at rest. In this case, the instability threshold and the characteristic time $\tau_g$ become time-dependent. The set-up time thus results from a trade-off between the slow relaxation of the viscosity and the increase of the critical acceleration and growth rate induced by shear-thickening. In order to get a more quantitative understanding of this phenomenon, a simple model is proposed thereafter.

\section{Model}
Our main assumption is that, once the surface waves have disappeared and the acceleration is set to $a$ at time $t=0$, the viscosity decreases exponentially with a characteristic time $\tau$: for $t>0$, $\eta(t)=\eta_0(1+\delta\exp(-t/\tau))$, where $\eta_0$ is the viscosity of the fluid at rest and $\delta$ accounts for shear-thickening induced during pre-shaking so that $\eta(t=0)=\eta_0(1+\delta)$. Moreover, it is known that, for a Newtonian fluid, $a_c\propto\eta$ and $\tau_g\propto\eta^{-1}$ \cite{Douady:1990}. If we assume our dilute micellar solution to be a Newtonian fluid whose only time-dependent parameter is the viscosity, we have:
\begin{eqnarray}
\label{ac.1}
a_c (t)&=&a_0 \frac{\eta (t)}{\eta_0}=a_0 (1+\delta e^{-t/\tau}),\\
\label{taug.1}
\tau_g (t)&=&\tau_{g0} \frac{\eta_0}{\eta (t)}=\frac{\tau_{g0}} {1+\delta e^{-t/\tau}},
\end{eqnarray}
where $\epsilon/\tau_{g0}$ is the growth rate for the fluid at rest.

For a given acceleration $a$ above $a_0$, two cases are possible: $a\leq a_c(0)$ or $a>a_c(0)$. If $a\leq a_c(0)$, the surface remains flat as long as $a<a_c(t)$. Let $t_1$ be the time needed for the viscosity to relax so that $a=a_c(t_1)$. Equation~(\ref{ac.1}) yields $t_1=\tau\ln(\delta/\epsilon)$. By changing the time origin $t \rightarrow t-t_1$, we are left with solving the case $a=a_c(0)$ {\it i.e.} $\epsilon=\delta$. For small amplitudes and time-dependent $\tau_g$ and $\epsilon$, eq.~(\ref{Ampl.0}) becomes 
\begin{equation}
\label{Ampl.3}
\tau_g (t) \frac{\hbox{\rm d}\xi}{\hbox{\rm d}t} = \left( \frac{a}{a_c(t)}-1 \right) \xi .
\end{equation}
Inserting eqs.~(\ref{ac.1}) and (\ref{taug.1}) and integrating eq.~(\ref{Ampl.3}) over time, one finds
$T_0/\epsilon=t_2-\tau (1-\exp(-t_2/\tau))$, where $t_2$ is the set-up time in the case $a=a_c(0)$ and $T_0=\tau_{g0}\ln(\xi_d/\xi_0)$. If one further assumes that $t_2/\tau \ll 1$, an expansion to second order in $t_2/\tau$ leads to $t_2=\sqrt{2T_0\tau/\epsilon}$.
Finally, when $a\leq a_c(0)$, the set-up time $t_a$ is the sum of the two terms $t_1$ and $t_2$: 
\begin{equation}
\label{taf.1}
t_a=t_1+t_2=\tau \ln \left( \frac{\delta}{\epsilon} \right) + \sqrt{\frac{2T_0\tau}{\epsilon}}.
\end{equation}
The first term corresponds to the decrease of $a_c(t)$ down to the imposed acceleration, while the second term derives from the growth of the surface wave amplitude up to $\xi_d$. 

Turning now to the case $a>a_c(0)$, one has $\epsilon(t) = (\epsilon-\delta e^{-t/\tau})/(\delta e^{-t/\tau}+1)$. Integrating eq.~(\ref{Ampl.3}) under the assumption $t_a/\tau \ll 1$ leads to
\begin{equation}
\label{taf.2}
t_a=\left( 1-\frac{\epsilon}{\delta} \right) \tau +\sqrt{\left( 1-\frac{\epsilon}{\delta} \right) ^2 \tau^2+\frac{2T_0\tau}{\delta}}.
\end{equation}
It is easily checked that for $\epsilon=\delta$, eqs.~(\ref{taf.1}) and~(\ref{taf.2}) give the same value for $t_a$ and that for $\epsilon\gg\delta$, the same scaling as for the fluid at rest, namely $t_a=T_0/\epsilon$, is recovered from eq.~(\ref{taf.2}).

Although our expression for $t_a$ involves four parameters, $a_0$, $T_0$, $\delta$, and $\tau$, there are actually only two free parameters $\delta$ and $\tau$. Indeed, as already mentioned, $a_0$ and $T_0$ are determined experimentally by fitting $t_a(\epsilon)$ for the fluid initially at rest using eq.~(\ref{ta.2}). Then $\delta$ and $\tau$ are found by fitting $t_a(\epsilon)$ for the pre-shaken fluid with eqs.~(\ref{taf.1}) and~(\ref{taf.2}), where $a_0$ and $T_0$ are fixed to the values found previously. The results of these fits are shown in fig.~\ref{f.3} and the various fit parameters are presented in table~\ref{t.1}. The fits are in good quantitative agreement with the experimental results. As expected, $a_0$ increases as $f$ increases and $\tau$ remains roughly constant $\tau\simeq 520$~s $\pm 10\%$. A posteriori, one can check that $t_2\ll \tau$ as soon as $\epsilon>0.002$.
\begin{table}	
	\begin{center}
		\begin{tabular}{|c|c|c|c|}
		\hline 
		$f$ (Hz)			&	65		&	85		&	135 	\\ \hline
		$a_0$ (m.s$^{-2}$)	&	7.3		&	10.9 	&	19.3	\\ \hline
		$T_0$ (s)			&	0.49	&	0.35	&	0.28 	\\ \hline
		$\tau$ (s)			&	501		&	565 	&	502		\\ \hline
		$\delta$ 			&	0.136	&	0.105	&	0.076 	\\ \hline
		\end{tabular}
	\end{center}
	\caption{\label{t.1} Best fit parameters used in eqs.~(\ref{taf.1}) and~(\ref{taf.2}) to model the $t_a$ vs $\epsilon$ data for $f=65$, $85$, and $135$~Hz. The resulting fits are shown in fig.~\ref{f.3}.}
\end{table}

Going back to the first experiment reported in fig.~\ref{f.2}(b), the same arguments can be used to understand the increase of $t_a$ with $t_v$ at a fixed $\epsilon$. Here we assume that during pre-shaking at $a_v=2a_0$, the viscosity increases with $t_v$ as $\eta(t_v)=\eta_0 (1+\delta_\infty (1-\exp(t_v/\tau)))$, where $\eta_0 (1+\delta_\infty)$ is the asymptotic viscosity and $\tau$ the characteristic time of shear-thickening. Note that, although the exponential decay proposed above is probably a reasonable model for the relaxation of the viscosity towards its value at rest, an exponential growth of $\eta(t)$ is clearly inconsistent with the rheological measurements shown in fig.~\ref{f.1}(a) which rather involve an induction time of about 200~s. In the present approach, we chose to keep an exponential form for $\eta(t)$ so that the equations remain tractable analytically. Following the path used to find eqs.~(\ref{taf.1}) and~(\ref{taf.2}), we end up with the following expressions: if $\delta_{\infty}(1-e^{-t_v/\tau})\geq\epsilon =0.05$,
\begin{equation}
\label{tab.1}
t_a = \sqrt{2\tau T_0 \left( 1+\frac{1}{\epsilon} \right) }+\tau \ln \left( \frac{\delta_{\infty}(1-e^{-t_v/\tau})}{\epsilon} \right) ,
\end{equation}
and if $\delta_{\infty}(1-e^{-t_v/\tau})<\epsilon=0.05$, 
\begin{equation}
\label{tab.2} 
t_a=\tau \left( 1-\frac{\epsilon}{\delta_{\infty}(1-e^{-t_v/\tau})} \right) +\sqrt{\tau^2 \left( 1-\frac{\epsilon}{\delta_{\infty}(1-e^{-t_v/\tau})}  \right) ^2+\frac{2T_0\tau}{\delta_{\infty}(1-e^{-t_v/\tau})}}.
\end{equation}
As shown in fig.~\ref{f.2}(b), the fit obtained with a characteristic time $\tau=461$~s and $\delta_{\infty}=0.149$ closely matches the experimental data. Such an accuracy of our crude exponential model for $\eta(t)$ points to a weak dependence of the set-up time on the details of $\eta(t)$.

\section{Discussion} 
Let us now discuss the present results and model. The various relaxation times $\tau$ found by fitting our simple model to the experimental data are strikingly similar to the characteristic times revealed in rheological experiments. We conclude that the same shear-thickening phenomenon is at play under pure shear flow and in the surface wave pattern induced by the Faraday instability. As in previous work under shear \cite{Boltenhagen:1997}, we tried to evidence turbid regions characteristic of a gel-like structure using light scattering. However such experiments did not reveal any significant increase of the turbidity neither locally nor globally. This is probably because the effect is too small or too localized compared to the one at constant shear rate. Indeed, as indicated by our model, the viscosity increase is much smaller under vibrations ($\delta\sim 10\%$) than under continuous shear ($\sim 1000\%$). We believe that the main reason for this discrepancy is that, in the Faraday instability, the shear rate at the surface oscillates in time and the period of the oscillation is much smaller than $\tau$ ($1/f=0.01\ll \tau\sim500$~s). Therefore, one should rather compare the shear strains $\tau \dot\gamma\simeq 2000$ under continuous shear to $2\pi \xi k_c\simeq 2$ in the Faraday experiment, where $\dot\gamma\simeq5$~s$^{-1}$ is the typical shear rate for shear-thickening and $\xi\simeq300$~$\mu$m is the amplitude of the surface waves at $\epsilon=1$. The fact that $\tau \dot\gamma \gg  2\pi\xi k_c$ could explain why the increase of viscosity is much smaller in the Faraday experiment but still noticeable through set-up time measurements. Finally, note that the comparison with pure shear flow (which involoves both rotation and elongation) may not be that relevant since the flow field induced by the Faraday instability is purely elongational besides a viscous sublayer \cite{Kityk:2006}. In the absence of any published data on the present micellar system under elongational flow, the issue of the exact interplay between Faraday waves and shear-thickening is left as an open question.

\section{Conclusion}
Set-up time measurements of the Faraday instability revealed a shear-thickening phenomenon induced by surface waves. A simple phenomenological model was proposed to account for the experimental observations. Even if no gel-like structure was directly evidenced, these experiments show that, far above the instability threshold, the surface deformation is large enough to induce a change in the microstructure of this particular complex fluid. Further experiments will deal with fluids whose shorter relaxation times may couple more strongly to the Faraday instability. An interesting question would then be whether the structuration of the flow into patterns with some characteristic wave number can be used to imprint some macroscopic spatial organization of the fluid microstructure with the same wave number.

\end{document}